 \newcommand{\theTitle}{Indistinguishability and improper mixtures}

 \newcommand{\theAbstract}{All quantum mixtures are what \dEsp has termed ``improper.''
His ``proper'' mixture cannot be created --- if \emph{welcher weg}, or distinguishing,
information exists, an improper mixture results, while in the absence of such
information, the resulting ``mixture'' is a pure state. \DEsp has claimed that an
interpretation of the improper mixture in terms of subensembles leads to logical
inconsistency; this claim is shown to be incorrect, as \dEsp's argument fails to account
for the indistinguishability of the pure-state subensembles.}

 \newcommand{\myname}{K. A. Kirkpatrick}
 \newcommand{\myaffiliation}{New Mexico Highlands University, Las Vegas, New Mexico 87701}
 \newcommand{\myemail}{\texttt{kirkpatrick@physics.nmhu.edu}}
  \documentclass[notitlepage,twoside,twocolumn,letterpaper,reqno,fleqn,10pt]{article}
 \usepackage[tbtags,nosumlimits]{amsmath}
 \usepackage{amssymb}
 \usepackage{QMarXiv} 
 \renewcommand{\versiondate}{\today}
 \usepackage{QM}      
 
 \arXiv 

 \ifReview%
  {\onecolumn}%
  {\setlength{\columnsep}{0.2in}
   \topmargin -0.8in
   \textheight 9.6in
   \oddsidemargin -0.325in
   \evensidemargin -0.425in
   \textwidth 7.25in}

 \newcommand{\PSI}[2]{\Psi%
     \def\xx{#1}\ifx\xx\empty{_{\,#2}}\else{{}^{\Sys{#1}}_{\,#2}}\fi}%
 \newcommand{\ketPsi}[2]{\ket{\PSI{#1}{#2}}\xspace}
 \newcommand{\ketPsiSM}{\ketPsi{\Sys{S\oplus M}}{}}
 \newcommand{\ketPsiSE}{\ketPsi{\Sys{S\oplus E}}{}}

\begin{document} 
 \doFront

\section{Introduction: Mixtures and their propriety}%
\VN\cite{vonNeumann55tr} introduced \emph{mixtures} of pure ensembles into quantum
mechanics exactly in the manner of classical probability, as a matter of ignorance.
Introducing the statistical operator (density matrix) as  the descriptor of a mixture, he
said (\RefCite{vonNeumann55tr}, p.~295) ``if we do not even know what state is actually
present --- for example, when several states $\phi_1,\,\phi_2,\dots$ with the respective
probabilities $w_1,\,w_2,\dots$ constitute the description,'' then the statistical
operator is $\bRho=\Sum{s}w_s\proj{\phi_{s}}$. (The state is pure, not mixed, if and only
if $\bRho$ is a projector: $\bRho=\bRho^2$.)

With the simple requirement that the statistics of a proposition not be changed by
conjunction with the trivial proposition in another system, \vN proved (p.~424) that the
unique statistical descriptor of a subsystem \SysS of a joint system \SysSM is given by
the partial trace: $\RhoS\DefEq\Trace[M]{\RhoSM}$. (This is pure only if
$\RhoSM=\RhoS\otimes\RhoM$ --- \ie only if \SysS is uncorrelated with its exterior.) This
statistical operator can always be expressed (in many ways) as a convex sum of pure-state
projectors, exactly the form of the ``ignorance'' mixture first introduced.

\subsection*{Propriety}%
{\DEsp}\cite{dEspagnat66,dEspagnat71b,dEspagnat76,dEspagnat90,dEspagnat95} has challenged
the validity of the trace-reduced statistical operator as the statistical descriptor of a
subsystem, introducing the category of ``propriety'' of mixtures:

The \emph{proper} mixture is simply a mixture of subensembles of pure states
\ket{\phi_{s}}, each with the weight $w_s$: $\RhoS=\Sum{s}w_s\proj{\phi_{s}}$ --- exactly
the ``ignorance'' mixture introduced by \vN.

The \emph{improper} mixture refers to the result of a trace-reduction of the statistical
operator \RhoSM of a composite system \SysSM:
$\Trace[M]{\RhoSM}=\Sum{s}w_s\proj{\phi_{s}}$.

\DEsp claims that an ignorance interpretation of the improper mixture is mathematically
inconsistent, and concludes that the improper mixture, the only candidate for the state
descriptor of a subsystem, is inadequate to this purpose. His proof consists in
reconstructing the state of a composite system, assumed pure, from the purported
subensembles arising from the trace-reductions, and obtaining the contradiction of a
mixed state.

However, the claimed contradiction is illusory. In \RefSec{S:argument}, I exhibit \dEsp's
argument and point out its error (which arises out of the neglect of
indistinguishability).

In \RefSec{S:MakeMixture}, I show that no physical process can create a \emph{proper}
mixture. Thus, it is the result of this paper that (i) all mixtures are ``improper,''
that is, are the state descriptors of subsystems, and (ii) no valid objection exists to
the use of statistical operators arising from trace-reduction.

\section{Rules of distinguishability}%
The effect of \emph{welcher weg} information was well-known and recognized from the very
earliest days of quantum mechanics.  Contrary to the early view that this effect is the
result of irreducible disturbance due to observation, it has become clear that it is an
intrinsic part of the formalism of quantum mechanics  (\cf, \eg
\RefCites{ScullySM78,ScullyWalther89,EnglertSW99b}). The clearest statement (and perhaps
most consistent use) of the principles regarding \emph{welcher weg} distinguishability is
that of Feynman.\cite{FeynmanVolIII}  Speaking \emph{propagator} language, Feynman said
(in paraphrase) ``To find the probability of a process: when the alternative processes
are indistinguishable, square the sum of their amplitudes; when distinguishable, sum the
squares of their amplitudes.''  Translating this into \emph{state} language, we say ``to
construct the appropriate state representative of the system: when the alternative states
are indistinguishable, add their vectors to get the state vector; when distinguishable,
add their projectors to get the state operator.'' (The vectors are weighted by the
amplitudes of the processes leading to the alternative states, the projectors by the
squares of these amplitudes).

\section{\DEsp's argument}\label{S:argument}%
\DEsp insists that the improper mixture, although represented by the same statistical
operator as the proper mixture, does not represent a mixture of subensembles in pure
states \set{\ket{\phi_{s}}}; the ignorance interpretation may not be applied to it. The
argument supporting this point is implicit in Feyerabend\cite{Feyerabend57}, is explicit
in \dEsp, and is frequently referred to in the literature of quantum interpretation. It
is particularly clearly stated in Hughes (\RefCite{Hughes89}, p.~150; see also p.~283),
whom I quote (taking some small liberties with the notation): \pagebreak
\begin{quote}
Consider a composite system in the \emph{pure state} \RhoSM, of which the component
states are the mixed states \RhoS and \RhoM. For the sake of argument, assume that
$\RhoS=a_1\proj{u_1}+a_2\proj{u_2}$, while $\RhoM=b_1\proj{v_1}+b_2\proj{v_2}$, with
$a_1\neq a_2$ and $b_1\neq b_2$, so there are no problems of degeneracy. Then, according
to the ignorance interpretation of \RhoS and \RhoM, system \SysS is really in one of the
pure states \ket{u_1} or \ket{u_2}, and system \SysM is really in one of the pure states
\ket{v_1} or \ket{v_2}. \dots\ But this would mean that the composite system is really in
one of the four states \ket{u_j\,v_k}, with probabilities $a_j b_k$ respectively
--- in other words, that the  composite system is in a \emph{mixed state}. Since this
contradicts our original assumption, the ignorance interpretation simply will not do.
\end{quote}

This argument is so clearly stated by Hughes that its error stands out: the claim that
``the composite system is in a \emph{mixed state}'' is not supportable --- nothing
\emph{external} to \SysSM distinguishes those states \ket{u_j\,v_k} from one another. We
must add the state vectors (not the projectors):
$\ketPsiSM=\Sum{jk}\psi_{jk}\ket{u_j\,v_k}$ --- a pure state. The contradiction does not
obtain; the argument fails to establish anything ``improper'' about these mixtures.

Hughes's presentation is very close to that of \RefCite{dEspagnat66} and of
\RefCite{dEspagnat71b}, p.~86.  In \RefCite{dEspagnat76}, p.~61, \dEsp's implicit
assumption that any combining of pure-state subensembles yields a mixture is even
clearer: he finishes with the explicit claim
\begin{quote}
\dots\ the fact that [the ensemble of the joint systems] should be describable as the
union of all the [pure-state subensembles] implies that it should be describable also by
a weighted sum of all the [pure-state projectors] \dots
\end{quote}
--- exactly, and incorrectly in this context, the rule for the combination of
distinguishable alternatives.

\section{The ``proper'' mixture cannot be created}\label{S:MakeMixture}%
How might we go about creating a mixture, in particular, a proper mixture?

Let us first exclude the obvious: It is always possible to destroy coherence (rather, its
observability), obtaining an apparent mixture, by sloppy technique --- the failure to
observe fringes with an interferometer on a wobbly table is not interesting! We assume
all relevant experimental technique to have been applied to protect the observability of
whatever coherence may be present --- there are no coherence-destroying temporal random
fluctuations in the coefficients. (\Cf \RefCite{EnglertSW99b}).

So, we return to \vN's original description of the mixed state (echoed by \dEsp for the
case of the proper mixture). The preparation of the system \SysS varies randomly among
the possible output states \set{\ket{\alpha_j}}; when \SysS is prepared in the state
\ket{\alpha_j}, the state of its relevant environment \Sys{E} (a system external to \SysS
such that \SysSE has no correlations with its exterior) is \ket{\eta_j}, and the
composite system \SysSE is described by the state $\ket{\alpha_j\,\eta_j}$. Because
\SysSE has no exterior correlations, these states are indistinguishable; the
Indistinguishability Rule requires the state of \SysSE to be pure, the sum
$\ketPsiSE=\Sum{s}\gamma_s\ket{\alpha_s\,\eta_s}$.

If the \set{\ket{\eta_j}} are all collinear, the state of \SysS is
$\ket{\Psi^{\SysS}}=\Sum{s}\gamma_s\ket{\alpha_{s}}$ --- pure, not a mixture! If the
\set{\ket{\eta_j}} are not all collinear, then, utilizing the Schmidt decomposition of
pure joint states, we have $\ketPsiSE=\Sum{s}\psi_s\ket{p_s\,a_s}$, with the
\set{\ket{p_j}} and the \set{\ket{a_j}} orthonormal, and more than one non-vanishing
coefficient $\psi_j$; then the state of \SysS is the \emph{improper} mixture
$\RhoS=\Sum{s}\abs{\psi_s}^2\proj{p_s}$. This analysis is exhaustive: it is not
physically possible to create \dEsp's proper mixture.

\subsection*{The Ancilla Theorem}%
If it is impossible to create a proper mixture, all physically existent mixtures must be
``improper'' --- correlated with another, external, system. Let us treat this important
matter formally.
\begin{theorem*}[Ancilla]%
If the state representative of a physical system \SysS is a mixture \RhoS, then there
must exist another physical system \Sys{M} such that, for every expression of \RhoS as a
convex sum of distinct (but not necessarily orthogonal) projectors
$\RhoS=\Sum{s}w_s\proj{\phi_{s}}$, there is a corresponding orthonormal set
\set{\ket{b_s}\in\HSM} in terms of which $\ketPsiSM=\Sum{s}\phi_s\ket{\phi_s\,b_s}$, with
$\abs{\phi_j}^2=w_j$.
\end{theorem*}
\Proof  Let \SysM be the relevant environment of \SysS; as discussed above, the state of
\SysSM must be pure: \ketPsiSM. Utilizing the Schmidt decomposition of pure joint states,
we have $\ketPsiSM=\Sum{s}\psi_s\ket{p_s\,a_s}$, with the \set{\ket{p_j}\in\HSS} and the
\set{\ket{a_j}\in\HSM} orthonormal. According to the GHJW Theorem\cite{Mermin99} (see
Appendix), \RhoS may be expressed in infinitely many ways as a convex sum of distinct,
but not necessarily orthogonal, projectors, all of which expressions --- and only such
expressions --- being generated from \ketPsiSE by the set of all unitary transformations
on \HS{E}: With the transformation $\ket{b_j}=\Sum{s}u_{js}\ket{a_s}$, we have
$\ketPsiSE=\Sum{s}\phi_s\ket{\phi_s\,b_s}$ and
$\RhoS=\Sum{s}\abs{\phi_s}^2\proj{\phi_s}$, where $\phi_j\ket{\phi_j}=\Sum{s}\psi_s
u^*_{js}\ket{p_s}$ and $\abs{\phi_j}^2=\Sum{s}\abs{\psi_j}^2\abs{u_{js}}^2$.
\mbox{}\hfill${\pmb\square}$~\vspace{-1ex}

\section{Discussion}%
The use of Hughes's form of the argument is not a straw man against \dEsp, whose argument
in \RefCites{dEspagnat66,dEspagnat71b,dEspagnat76} is essentially the same. In the more
recent \RefCites{dEspagnat90} and \onlinecite{dEspagnat95}, \dEsp ``simultaneously
measures'' the two systems to build up a mixture. But measurement is hardly an innocent
activity in quantum mechanics: it creates the very distinguishability that does, in fact,
produce the mixed state of the conclusion; in doing so, modifies the state of the joint
system so it may no longer be assumed pure. Again, there is no contradiction.

For a recent discussion utilizing d'Espagnat's distinction of propriety, consider
\RefCite{VermaasDieks95}, which gives as the ``paradigmatic example'' of a proper mixture
the case of a preparation which depends on the result of a coin toss. But such a
preparation does \emph{not} lead to a proper mixture: The coin and the system are
entangled; the system's state is indeed a mixture, but improper --- it is obtained by a
partial trace over the coin's space.

The error in \dEsp's argument has ``caused some confusion in the theory of measurement,''
in Jammer's phrase (\RefCite{Jammer74}, pp.~479-80, footnote), referring to the argument
of \RefCites{dEspagnat66} and \onlinecite{dEspagnat76}) --- but in the sense opposite to
that meant by Jammer. The ``confusion'' referred to is the repeated rediscovery of what
might be called the ``Landau-collapse'' analysis of quantum measurement
(\RefCite{Landau27}), in which tracing out the measuring apparatus leaves the system in
the appropriate mixture, obviating \vN's Process--1 collapse. Jammer implies that \dEsp's
argument has invalidated this understanding. But it is \dEsp's argument which is
incorrect; indeed, the subensembles described by the mixture are observable through their
correlations with the ancillary system. The post-measurement \RhoS mixture --- exactly
the result \vN requires (\RefCite{vonNeumann55tr}, p.~347) at the start of the argument
which ends with the introduction of Process--1 (p.~351) --- arises naturally, without use
of Process--1.

\subsection*{Related work of \dEsp}%
As the title of \RefCite{dEspagnat95} would tell us, much of \dEsp's interest surrounding
this issue, and quantum mechanics in general, involves the issue of the determinacy of
the values of variables: ``realism.'' Because I do not even mention this issue, much less
discuss it at length, my criticism of his argument may seem rather ``schematic''%
\footnote{Professor \dEsp's characterization (private communication).} %
compared with \dEsp's discussions. But the application of the distinguishability
heuristics to \dEsp's ensemble argument requires no stance regarding the determinacy of
variables. (Furthermore, as Examples 3 and 4 of \RefCite{KAK:ClassicalExamples}
illustrate, indeterminate-valued variables occur naturally in irreducibly
nondeterministic systems, systems which are perfectly ordinary, real, and entirely
distinct from quantum mechanics. Value-indeterminacy is not a problem for the
interpretation of quantum mechanics just so long as quantum mechanics is taken to be
irreducibly nondeterministic.)

In a footnote on p.~1154 of \RefCite{dEspagnat90}, responding to another's charge that
the proper mixture does not exist, \dEsp claims that if the concept of a proper mixture
is void, then the statement ``immediately after an observable has been measured it
\emph{has} the observed value'' is also a void statement, if the expression ``it
\emph{has}'' is to have its ordinary, commonsense meaning. I must disagree with \dEsp's
claim. Whatever one might mean by the ``measurement of an observable,'' it must minimally
involve the correlation of the values of that observable with something outside the
system. It follows directly, from basic rules of quantum mechanics, that, after this
first stage of a measurement, the state of the system itself is an \emph{improper}
mixture. \DEsp's proper mixture (whatever that might be) does not arise in measurement
--- the issue of its existence cannot therefore be the basis for any argument (physical
or metaphysical) regarding measurement. This same error is seen in Section 3 of
\RefCite{dEspagnat98b}, where \dEsp considers a standard measurement process correlating
the system with the pointer.  He describes ``the ensemble $\EnsSym$ of all the pointers
\dots\ composed of $\mu$ subensembles $\EnsSym_1,\cdots,\EnsSym_\alpha,
\cdots,\EnsSym_\mu$, the components of each $\EnsSym_\alpha$ being pointers really lying
in one definite interval, the one labeled $\alpha$.'' \DEsp then argues ``since $\EnsSym$
must be the addition of all the $\EnsSym_\alpha$, it is \dots\ a `proper mixture'.'' But
each element in $\EnsSym_\alpha$ is correlated with a corresponding element of the
measured system, causing $\EnsSym$ to be an \emph{improper} mixture. (Curiously, earlier
in that section \dEsp stated that $\EnsSym$'s density matrix was obtained by partial
tracing the system-plus-instrument density matrix --- an improper mixture by, it would
seem, his own standard.)

In the last paragraph of Section 3 of \RefCite{dEspagnat98b}, \dEsp considers  systems
``that cannot be distinguished from one another by any measurement,'' saying that, taking
a realistic approach, ``there is no reason why we should not consider ensembles of
systems of the same type lying in different quantum states,'' and claims that ``there are
cogent reasons to describe such ensembles by density matrices and call them mixtures.''
But, as argued in \RefSec{S:MakeMixture} above, if these systems indeed cannot be
distinguished, then an ensemble of them will be described by a pure state operator ---
not a mixture, proper or improper.

\subsection*{The ``ignorance'' interpretation of mixtures}%
That all mixtures are improper, thus correlated with ancillae, gives a clearer
understanding of the temptation to their interpretation by ignorance. A mixed state of
\SysS, $\bRho=\Sum{s}w_s\proj{\Phi_s}$, implies a correlation with an orthonormal basis
(equivalently, with the values of some variable) of a system in the world exterior to
\SysS; thus we may identify, with certainty, the state \ket{\Phi_j} of each occurrent
\SysS without interacting with \SysS.

The fact (as shown by the GHJW Theorem) that \emph{every} variable in the ancillary
system generates a distinct expansion of $\bRho$ as a convex sum of projectors does not
introduce a problem with uniqueness: these expansions of $\bRho$ are all equal. Further,
exactly \emph{one} such ancillary variable may, in the occurrent actuality, be observed,
thus exactly \emph{one} corresponding filter may be applied; there is no possibility of
incompatible state assignment.

\section{Conclusion}%
\DEsp's argument --- concluding that it is inconsistent to consider as a mixture the
state obtained by the trace-reduction of entangled systems --- fails by failing to take
into account the issue of indistinguishability. Further, fundamental rules of quantum
mechanics prohibit the production of that which \dEsp calls a ``proper'' mixture. The
category \emph{proper mixture} is physically empty; states are either ``improper''
(correlated with ancillae) or pure (completely uncorrelated with the exterior).

The failure of \dEsp's argument removes the only objection of substance (rather than of
interpretive faith) to the Landau-collapse view of measurement decoherence.


\appendix
\renewcommand{\theequation}{A\arabic{equation}}%
\setcounter{equation}{0}
\renewcommand{\theTheorem}{A\arabic{Theorem}}
\setcounter{Theorem}{0} \setcounter{Definition}{0}

\section*{Appendix. The GHJW Theorem}%
Here is a concise presentation of a theorem originally due to Gisin\cite{Gisin89} and
Hughston, Josza and Wootters\cite{HughstonJW93}, with proof simplified by Mermin and
others\cite{Mermin99}; Mermin calls it the GHJW Theorem.

\begin{lemma*}
\ket{\Psi} and \ket{\Phi} are vectors in \HAB. If $\TrB{\proj{\Psi}}=\TrB{\proj{\Phi}}$,
then there exists a unitary transformation \U on \HB such that
$\ket{\Psi}=\big(\One\otimes\U\big)\ket{\Phi}$.
\end{lemma*}
\begin{proof} Since it is positive and Hermitian,
$\TrB{\proj{\Psi}}=\Sum{j}{w_j\,\proj{p_j}}$, $w_j>0$ and \set{\ket{p_j}\in\HA}
orthonormal. Then, for any complete and orthonormal set \set{\ket{a_k}\in\HB},
$\ket{\Psi}=\Sum{jk}{\psi_{jk}\ket{p_j\,a_k}}$. Introducing the un-normalized kets
$\ket{\beta_j}\DefEq\Sum{k}{\psi_{jk}\ket{a_k}}$, we have
$\ket{\Psi}=\Sum{j}{\ket{p_j\,\beta_j}}$. Then
$\Sum{j}{w_j\,\proj{p_j}}=%
 \Sum{jj'}{\braket{\beta_{j}}{\beta_{j'}}\ket{p_{j'}}\bra{p_{j}}}$;
thus $\braket{\beta_{j}}{\beta_{j'}}=w_j\KDelta{j}{j'}$, and we can write
$\ket{\beta_j}=\sqrt{w_j}\ket{b_j}$, the \set{\ket{b_j}} orthonormal; hence
$\ket{\Psi}=\Sum{j}{\sqrt{w_j}\,\ket{p_j\,b_j}}$. The same argument leads to
$\ket{\Phi}=\Sum{j}{\sqrt{w_j}\,\ket{p_j\,c_j}}$. The sets \set{\ket{b_j}} and
\set{\ket{c_j}} extended to orthonormal bases of \HB are related by a unitary
transformation $\ket{b_j}=\U\ket{c_j}$ (explicitly, $\U=\Sum{s}\ket{b_s}\bra{c_s}$).
\end{proof}

\begin{theorem*}[GHJW]
\ket{\Psi} is a vector in \HAB; $\rhoA=\TrB{\proj{\Psi}}$. For any expression of $\rhoA$
in the form of a convex sum of distinct 1-projectors on \HA,
$\rhoA=\sum_{j=1}^{m}{f_j\proj{\phi_j}}$, $m\leq\text{dim}\,\HB$, there exists an
orthonormal set \set{\ket{c_j}} in \HB such that\linebreak
$\ket{\Psi}=\sum_{j=1}^{m}\sqrt{f_j}\ket{\phi_j}\otimes\ket{c_j}$.
\end{theorem*}
\begin{proof} Construct
$\ket{\Psi'}=\sum_{j=1}^{m}\sqrt{f_j}\ket{\phi_j}\otimes\ket{d_j}$, with \set{\ket{d_j}}
an arbitrary set of $m$ orthonormal vectors in \HB. By the Lemma, there exists a unitary
transform \U on \HB such that
$\ket{\Psi}=\big(\One\otimes\U\big)\ket{\Psi'}=%
\sum_{j=1}^{m}\sqrt{f_j}\ket{\phi_j}\otimes\U\ket{d_j}$; $\ket{c_j}\DefEq\U\ket{d_j}$.
\end{proof}


\small\providecommand{\bysame}{\leavevmode\hbox to3em{\hrulefill}\thinspace}

\end{document}